\documentclass[
reprint,
superscriptaddress,
amsmath,
amssymb,
aps,
longbibliography,
pra,
showpacs,
floatfix
]{revtex4-2}
\usepackage{hyperref}
\usepackage{graphicx}
\usepackage{xcolor} 

\usepackage{tabularx}
\usepackage{multirow}
\usepackage{dcolumn} 
\newcolumntype{d}[1]{D{.}{.}{#1}}
\newcolumntype{L}[1]{>{\raggedright\arraybackslash}p{#1}}
\newcolumntype{C}[1]{>{\centering\arraybackslash}p{#1}}
\newcolumntype{R}[1]{>{\raggedleft\arraybackslash}p{#1}}

\usepackage{braket}

\usepackage{comment}
\usepackage{enumitem}
\usepackage{soul}

\hypersetup{colorlinks=true, 
    linkcolor={blue},
    citecolor={blue}, 
    urlcolor={blue}
}

\begin{document}

\title{Multicolor phonon excitation in terahertz cavities}

\author{Omer Yaniv}
\email{yaniv@mail.tau.ac.il}
\affiliation{School of Physics and Astronomy, Tel Aviv University, Tel Aviv 6997801, Israel}

\author{Dominik~M.\ Juraschek}
\email{d.m.juraschek@tue.nl}
\affiliation{School of Physics and Astronomy, Tel Aviv University, Tel Aviv 6997801, Israel}
\affiliation{Department of Applied Physics and Science Education, Eindhoven University of Technology, 5612 AP Eindhoven, Netherlands}

\date{\today}


\begin{abstract}
Driving materials using light with more than one frequency component is an emerging technique, enabled by advanced pulse-shaping capabilities in recent years. Here, we translate this technique to lattice vibrations, by exciting multicolor phonons using terahertz cavities. In contrast to light, phonon frequencies are determined by the crystal structure and cannot readily be changed. We overcome this problem by tuning the frequencies of phonon polaritons in terahertz cavities to achieve the desired frequency ratios necessary for phononic Lissajous figures. This methodology enables dynamical crystallographic symmetry breaking and the creation of staggered phonon angular momentum and magnetic moment patterns.

\end{abstract}

\maketitle


\section{Introduction}

Exciting lattice vibrations with light provides a unique tool for controlling the properties of materials by inducing changes in their crystal structure. This technique has enabled ultrafast control of various electronic phases in recent years, including ferroelectricity \cite{Nova2019,Li2019,Henstridge2022,Kwaaitaal2023}, magnetism \cite{Disa2020,Afanasiev2021,Stupakiewicz2021,Disa2023}, and superconductivity \cite{Mitrano2016,Fava2024_superconductivity}, mediated by interactions between different vibrational modes (phonons) in the crystal. When driven with circularly polarized light, chiral phonons carrying angular momentum can further be generated that produce effective magnetic fields on the atomic scale \cite{nova:2017,juraschek2:2017,Juraschek2019,Geilhufe2021,Juraschek2022_giantphonomag,Luo2023,Basini2024,Davies2024}. The vibrational motion of the atoms in the crystal hereby acts as a periodic drive for the electronic system that can be captured within Floquet theory \cite{Shin2018,Hubener2018,Chaudhary2020_phononFloquet}, analogously to the well-established Floquet driving of electronic bands with light \cite{Oka2009,Wang2013_Floquet,Oka2019}.

At the same time, Floquet engineering of materials has evolved beyond simple periodic drives and more complex excitation schemes involving light with more than one frequency component, also called ``multicolor'' or ``two-tone'' drives, promise advanced control over materials properties \cite{Sandholzer2022,Castro2022,Wang2023_multicolor,Strobel2023,Murakami2023,Chen2024_multicolor,Chen2024_multicolor2,Pena2024}. While the frequencies of light are tunable with modern techniques, the frequencies of phonons are determined by the structure and atomic composition of the crystal and cannot arbitrarily be changed. Accordingly, multicolor phonon driving has remained elusive as a useful tool for dynamical materials control.

Here, we theoretically demonstrate a methodology to achieve multicolor phonon excitation by tuning the vibrational frequencies through the formation of cavity-phonon polaritons in terahertz cavities. We show that steady-state phononic Lissajous figures can be generated when two-tone drives are tuned into a $\omega:2\omega$ frequency ratio. These phononic Lissajous figures enable dynamical crystallographic symmetry breaking without requiring nonlinear phonon interactions and further produce staggered angular momenta and magnetic fields that can be used to control spatially varying magnetic ordering, such as in antiferromagnets or spin spirals.


\section{Cavity-phonon polariton splitting}
\begin{figure}[t]
\centering
\includegraphics[scale=0.3]{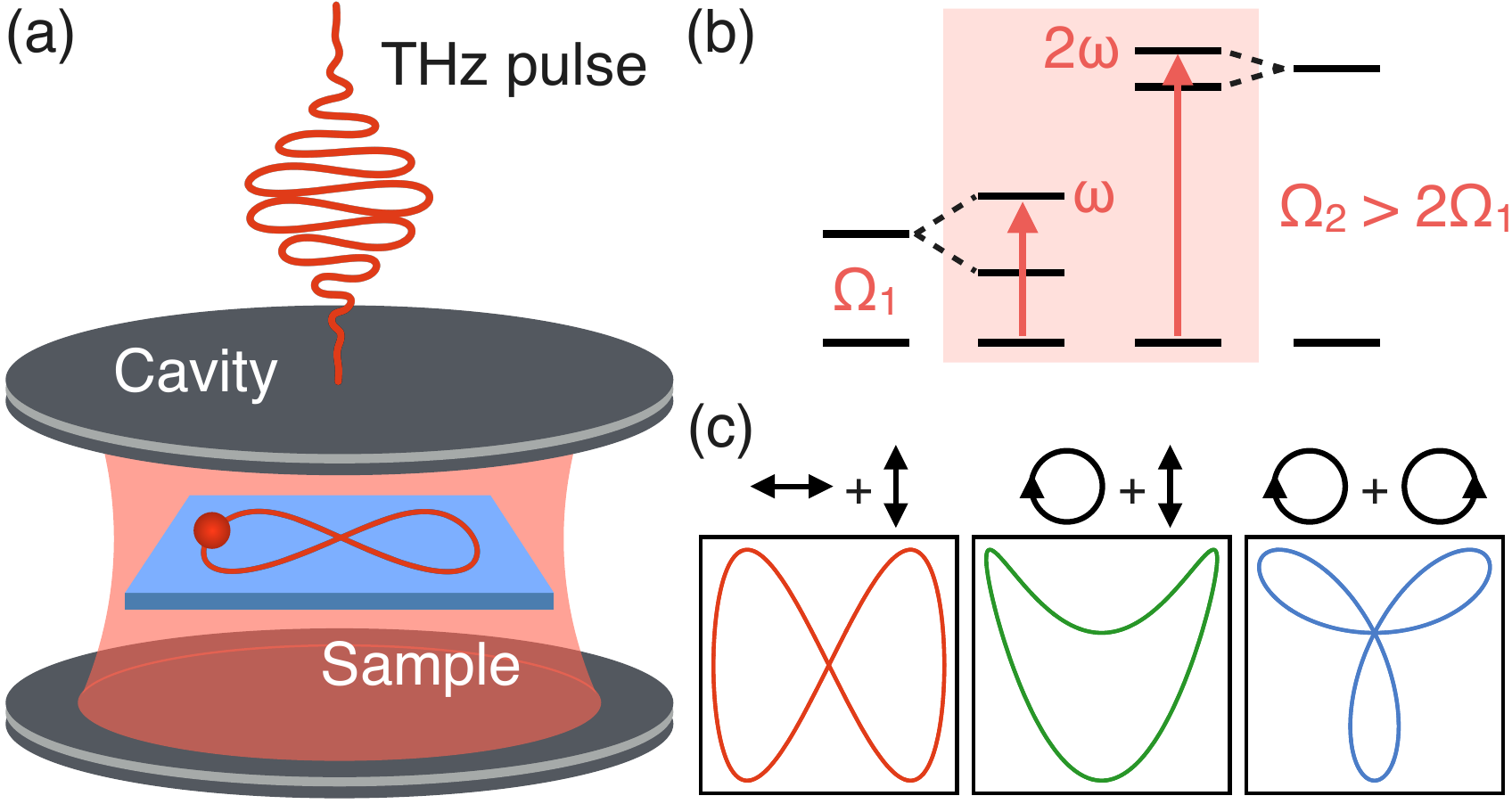}
\caption{
\label{fig:cavity}
{\small
\textbf{Cavity-induced multicolor phonons.} (a) A multicolor terahertz pulse drives an optical cavity, which in turn drives multicolor phonons in a material. (b) The cavity-phonon polariton splitting of two infrared-active modes, $\Omega_1$ and $\Omega_2$, can be tuned by the cavity. Here, the two upper polariton branches are tuned into an $\omega : 2\omega$ ratio and excited by the pulse. (c) Lissajous figures created by different combinations of linearly and circularly polarized phonons: eight curve, arrowhead, and cloverleaf.
}
}
\end{figure}

We begin by introducing the concept of infrared (IR)-active phonons coupled to an optical cavity, as illustrated in Fig.~\ref{fig:cavity}(a). In this setup, a slab of a material exhibiting IR-active phonons is placed in the center of a Fabry-Perot cavity formed by two parallel mirrors. The interaction between the IR-active phonons and the cavity leads to the formation of cavity-phonon polaritons \cite{Sentef2018,Ashida2020,Jarc2022} which can be excited by resonantly pumping the cavity with an external terahertz pulse \cite{Juraschek2021_4}. Because the phonon and cavity modes are required to couple resonantly, only the fundamental cavity mode is relevant in this process. To describe the interaction between the fundamental cavity mode, the infrared (IR)-active  phonon modes, and the electric field of the external pulse, we follow the formalism previously established in Ref.~\cite{Juraschek2021_4}. The coupled equations of motion for the cavity-phonon polariton dynamics are given by
\begin{align}
\ddot{A}_{i} + \kappa_c \dot{A}_{i} + \omega_c^2 A_{i} & = B E_{i}(t) + D \ddot{Q}_{\mathrm{\nu},i}, \label{eq:A1} \\
\ddot{Q}_{\mathrm{\nu},i} + \kappa_\mathrm{\nu} \dot{Q}_{\mathrm{\nu},i} + \Omega_{\mathrm{\nu}}^{2}Q_{\mathrm{\nu},i} & = G A_{i}, \label{eq:IR}
\end{align}
where $A_{i}$ represents the amplitude of the fundamental cavity mode along spatial direction $i$, and $Q_{\nu,i}$ denotes the normal mode coordinate (amplitude) of an IR-active phonon mode $\nu$ polarized along $i$, expressed in units of pm$\sqrt{u}$, where $u$ is the atomic mass unit. Spatial directions $i=a,b$ are defined with respect to the crystal axes of the sample. $\Omega_{\nu}$ is the phonon eigenfrequency and $\kappa_{\nu}$ is the phonon linewidth. Respectively, $\omega_c = c\pi/L$ is the frequency of the fundamental cavity mode, where $L$ is the length of the cavity, and \( \kappa_c \) is its linewidth. \( E_i(t) \) is the electric field of the external terahertz pulse. The interaction between the cavity mode and the IR-active phonon is described by two terms: The first term describes the coupling of the phonon to the cavity electric field via its mode effective charge, $Z_\nu$, through the coupling coefficient  \(G = \sin(k z_0) Z_{\mathrm{\nu}}\), where \( z_0 \) denotes the sample position at the center of the Fabry-Perot cavity and \( k \) is the wave vector of the fundamental cavity mode. 
The second term accounts for the phonon back action on the cavity mode, characterized by  \(D = -2Z_{\mathrm{\nu}} \Delta z \sin(k z_0)/(V_{\rm c} \varepsilon_0 L)\), where \( \varepsilon_0 \) is the vacuum permittivity and $V_c$ the unit-cell volume. Interaction of the cavity with the external field is given by \( B = 2\omega_c \sqrt{\omega_c \kappa_c/\pi} \).

As illustrated schematically in Fig.~\ref{fig:cavity}(b), the cavity interacts with the electric dipole moment of the IR-active phonon, resulting in the formation of two cavity-phonon polariton branches, where the frequencies of the upper and lower branches are determined by the cavity frequency. Accordingly, the cavity-phonon polariton splitting can be controlled by the length of the cavity. We can exploit this tuning capability to modify the frequencies of the cavity-phonon polariton branches in order to achieve a $\omega : 2\omega$ frequency ratio required to form the phononic Lissajous figures. If the $\omega$ and $2\omega$ phonon-polariton branches arise from degenerate phonon modes, exciting them with varying degrees of polarity is possible. Specifically, both phonons can be excited with linear, circular, or mixed circular-linear polarizations, resulting in the prototypical Lissajous curves illustrated in Fig.~\ref{fig:cavity}(c). 

\begin{figure}[t]
\centering
\includegraphics[scale=0.275]{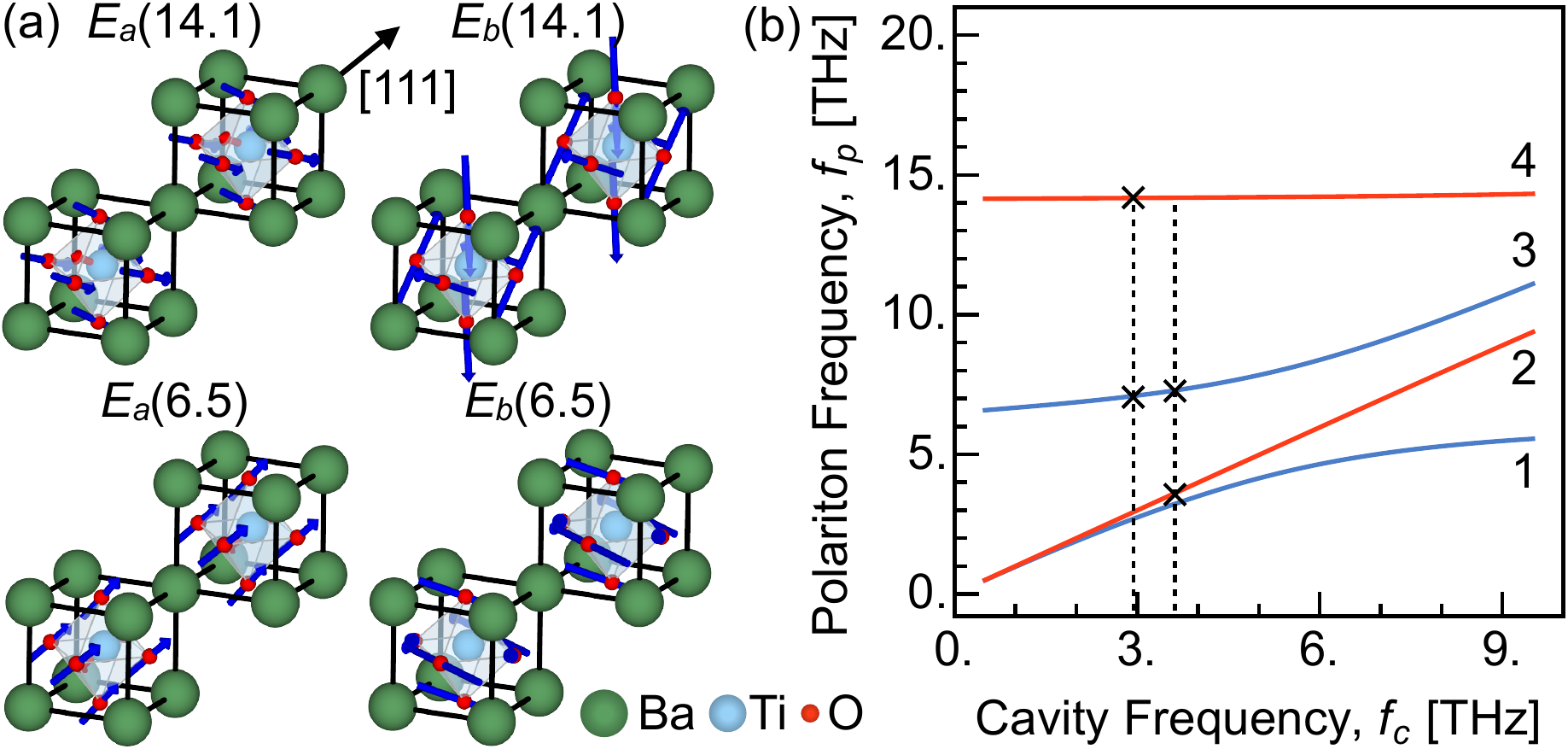}{

\caption{\small
\textbf{Cavity-phonon polaritons.} (a) Atomic displacements in BaTiO$_3$ corresponding to the doubly degenerate $E$(6.5) and $E$(14.1) modes. (b) Polariton dispersion as a function of fundamental cavity frequency, showing upper and lower branches for the $E$(14.1) mode in red and those for the $E$(6.5) mode in blue. Crosses and dashed lines mark the cavity frequency and polariton branches at which an $\omega:2\omega$ frequency ratio is achieved, $\omega_c=3.66$~THz for branches 2 and 3, and $\omega_c=2.97$~THz for branches 3 and 4.
}
\label{fig:polaritons}
}

\end{figure}

We demonstrate this methodology at the example of barium titanate (BaTiO$_3$) in its low-temperature rhombohedral phase with point group $C_{3v}$. BaTiO$_3$ is a ferroelectric insulator with a band gap of more than 3~eV \cite{evarestov2012first}, exhibiting IR-active phonon modes with large electric dipole moments, which makes it suitable for phonon driving. We are interested in the doubly degenerate phonon modes with $E$ irreducible representations and polarizations in the $ab$-plane of the crystal. The 6.5~THz and 14.1~THz phonon modes, whose $\omega:2\omega$ ratio is off by 8~percent, are appropriate candidates for demonstrating the mechanism. 

We calculated the phonon eigenfrequencies, eigenvectors, and the Born effective charge tensors using the density functional theory formalism as implemented in VASP \cite{kresse:1996,kresse2:1996}, and the frozen-phonon method as implemented in phonopy \cite{Togo2015}. We used the default projector augmented wave (PAW) pseudopotentials for each atom and converged the Hellmann-Feynman forces to 50~$\mu$eV/\AA. We used a plane-wave energy cut-off of 700~eV and a 8$\times$8$\times$8 k-point gamma-centered Monkhorst-Pack mesh to sample the Brillouin zone \cite{Monkhorst/Pack:1976}. For the exchange-correlation functional, we chose the PBEsol form of the generalized gradient approximation (GGA) \cite{csonka:2009}. We find the mode effective charges to be $Z_\nu=2.23$~$e/{\sqrt{u}}$ for the $E$(14.1) mode and $Z_\nu=0.82$~$e/{\sqrt{u}}$ for the $E$(6.5) mode, respectively. We further use phenomenological values for the linewidths of the phonon modes and the fundamental cavity mode of 5\%{} of the respective eigenfrequencies.

In Fig.~\ref{fig:polaritons}, we show the formation of cavity-phonon polariton branches in BaTiO$_3$. We visualize the phonon displacement patterns in Fig.~\ref{fig:polaritons}(a), which involve primarily displacements of the oxygen ions. The frequencies of the cavity-phonon polariton branches are obtained by  solving Eqs.~\eqref{eq:A1} and \eqref{eq:IR} without an external field, $E_i(t) = 0$, in Fourier space (see Supplemental Material for details).
In Fig.~\ref{fig:polaritons}(b), we show the polariton frequencies as a function of the fundamental cavity frequency. The four phonon polariton branches correspond to the coupling of the cavity mode to the $E(6.5)$ mode (red curves) and to the coupling to the $E(14.1)$ mode (blue curves). 
 
\section{Multicolor phonon dynamics}


\begin{table}[b]
\centering
\caption{Values of \( E_{\alpha,i} \) (\(\mathrm{MV}/\mathrm{cm}\)) and \( \phi_{\alpha,i} \) for the three polarization configurations linear (LL), mixed circular-linear (CL), and circular (CC) used in Figure 3 and Figure 4.}
\begin{tabular}{ccccccc}
\hline\hline
\multirow{2}{*}{Parameters} & \multicolumn{3}{c}{Figure 3} & \multicolumn{3}{c}{Figure 4} \\
\cline{2-7}
 & LL & CL & CC & LL & CL & CC \\
\hline
\( E_{2,a} \), \( \phi_{2,a} \) & 5, 0 & 5, 0 & 5, $\frac{\pi}{2}$ & 5, 0 & 5, 0 & 5, $\frac{\pi}{2}$ \\
\( E_{2,b} \), \( \phi_{2,b} \) & 0, 0 &5, $\frac{\pi}{2}$ & 5,0 & 0, 0 & 5, $\frac{\pi}{2}$ & 5, 0 \\
\( E_{3,a} \), \( \phi_{3,a} \) & 0, 0 & 0, 0 & 0.95, $\frac{\pi}{2}$ & 0, 0 & 0, 0 & 0.95, $\frac{\pi}{2}$ \\
\( E_{3,b} \), \( \phi_{3,b} \) & 0.95, $\frac{\pi}{2}$ & 0.95, 0 & 0.95, 0 & 0.95, 0 & 0.95, 0 & 0.95, 0 \\
\hline\hline
\end{tabular}
\label{tab:merged_polarization}
\end{table}

\begin{figure*}[t]
    \centering
    \includegraphics[scale=0.5225]{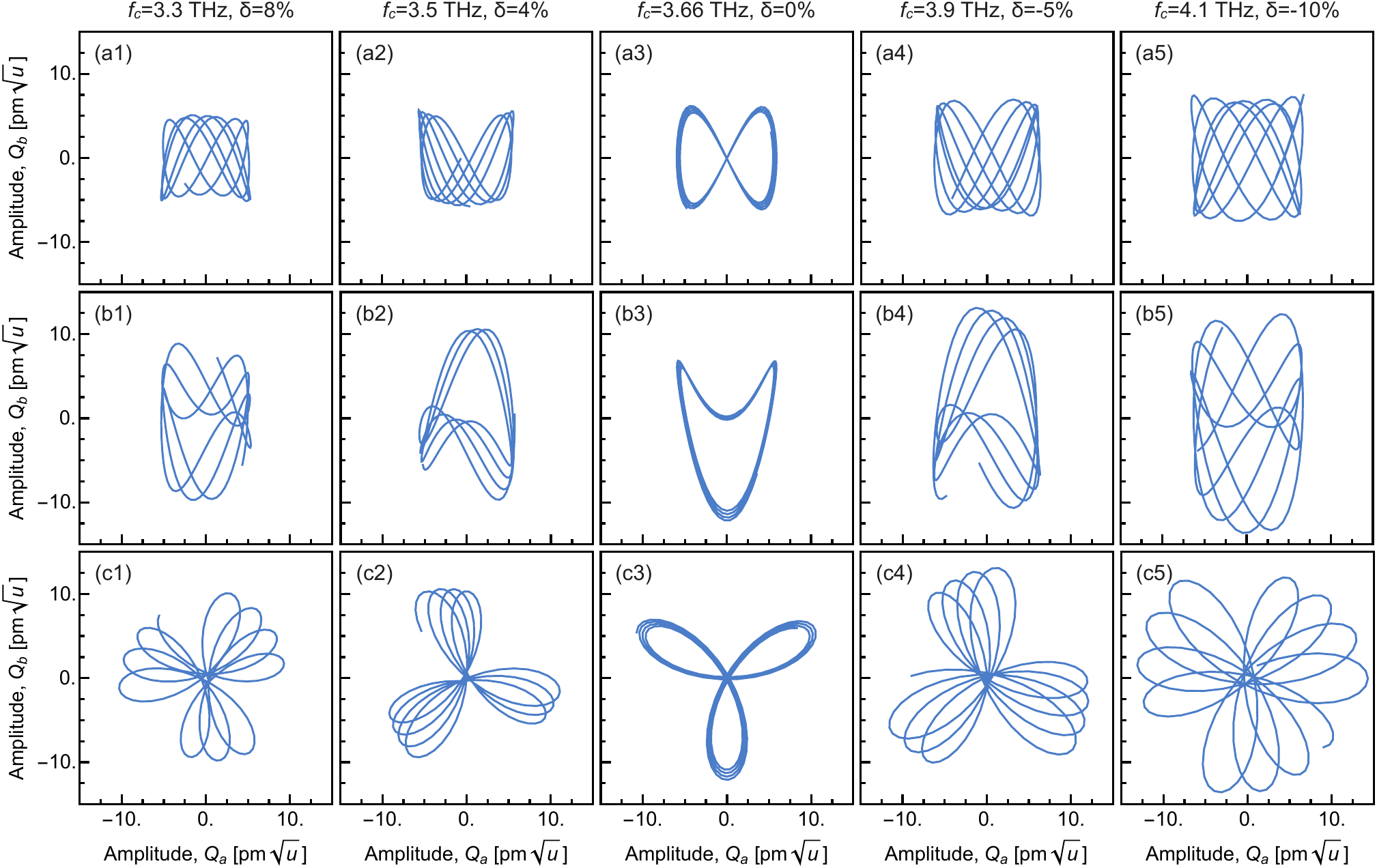}
    \caption{\small
        \textbf{Phononic Lissajous figures.} Time evolution of the phonon trajectories in the $ab$ plane of the crystal for the time interval of 1~ps to 2~ps after excitation. We show the dynamics for five different cavity frequencies, tuning the cavity-phonon polariton frequencies into and out of the $\omega:2\omega$ frequency ratio, reached at $\omega_c=3.66$~THz and yielding polariton frequencies of $\Omega_3=7.29$~THz and $\Omega_2=3.65$~THz. When the cavity frequency satisfies the  $\omega:2\omega$  ratio, the Lissajous figures exhibit closed trajectories, leading to stable crystallographic symmetry breaking. As the cavity frequency is detuned, the phononic trajectories progressively deviate from closed orbits, disrupting the symmetry-breaking mechanism. (a1-a5) Dynamics for linear excitations of the polariton branches, creating an eight curve when reaching the ideal frequency ratio. (b1-b5) Dynamics for mixed circular-linear excitations, creating an arrowhead when reaching the ideal frequency ratio. (c1-c5) Dynamics for circular excitations, creating a cloverleaf curve when reaching the ideal frequency ratio.}

    \label{fig:Lissajous}
\end{figure*}

We now demonstrate the generation of the basic phononic Lissajous figures outlined in Fig.~\ref{fig:cavity}(c), by driving the cavity-phonon polariton branches at the $\omega:2\omega$ frequency ratio. We choose the upper branch of the $E(6.5)$ polariton (no.~3) at 7.29~THz and the lower branch of the $E(14.1)$ polariton (no.~2) at 3.65~THz. The corresponding dynamics for branches 4 and 3 are shown in the Supplemental Material. We solve Eqs.~\eqref{eq:A1} and \eqref{eq:IR} numerically for varying cavity frequencies around 3.66 THz to illustrate how the phononic Lissajous figures can be generated by tuning into and out of the $\omega:2\omega$ frequency ratio. We model the electric field of the multicolor terahertz pulse as 
\begin{equation}
E_i(t) =  e^{\frac{-t^2}{2 (\tau/\sqrt{8 \ln 2})^2}}  
\sum_{\alpha = 2,3}  E_{\alpha,i} \cos(\Omega_\alpha - \phi_{\alpha,i} ),
\end{equation}
where the pulse contains two center frequency components, $\Omega_{\alpha}$, tuned into resonance with the frequencies of the respective polariton branches, $\Omega_2$ and $\Omega_3$. $E_{\alpha,i}$ is the peak electric field, which we adjust for each of the branches and spatial directions individually. The carrier envelope phase, $\phi_{\alpha,i}$, controls the polarization configuration (linear, circular, mixed circular-linear). $\tau$ is the full width at half maximum (FWHM) pulse duration.

In Fig.~\ref{fig:Lissajous} we show the phonon dynamics induced by an ultrashort terahertz pulse with a FWHM duration of 1~ps and peak electric fields of 0.95~MV/cm and 5~MV/cm for branches 2 and 3, respectively. We present the Lissajous figures created by the phonon trajectories in the $ab$ plane of the crystal for the interval between 1~ps to 2~ps after the excitation. The trajectories are shown for varying cavity frequencies presented from left to right and for each one of the three pulse polarizations (linear, mixed circular-linear, and circular) presented from top to bottom. Each of the cavity frequencies is associated with a detuning fraction defined as $\delta=1-2\Omega_2/\Omega_3$. When the cavity frequency is tuned precisely to obtain the $\omega:2\omega$ frequency ratio ($\delta=0$, $\Omega_2=3.65$~THz, $\Omega_3=7.29$~THz), the generated Lissajous figures exhibit closed trajectories that enable steady-state crystallographic symmetry breaking. Detuning the cavity frequency in turn leads away from the $\omega:2\omega$ frequency ratio and quickly dephases the Lissajous figures, no longer preserving their characteristic shapes and indicating that precise frequency control is critical for maintaining the desired phononic trajectories. 

\begin{figure}[t]
\centering
\includegraphics[scale=0.39]{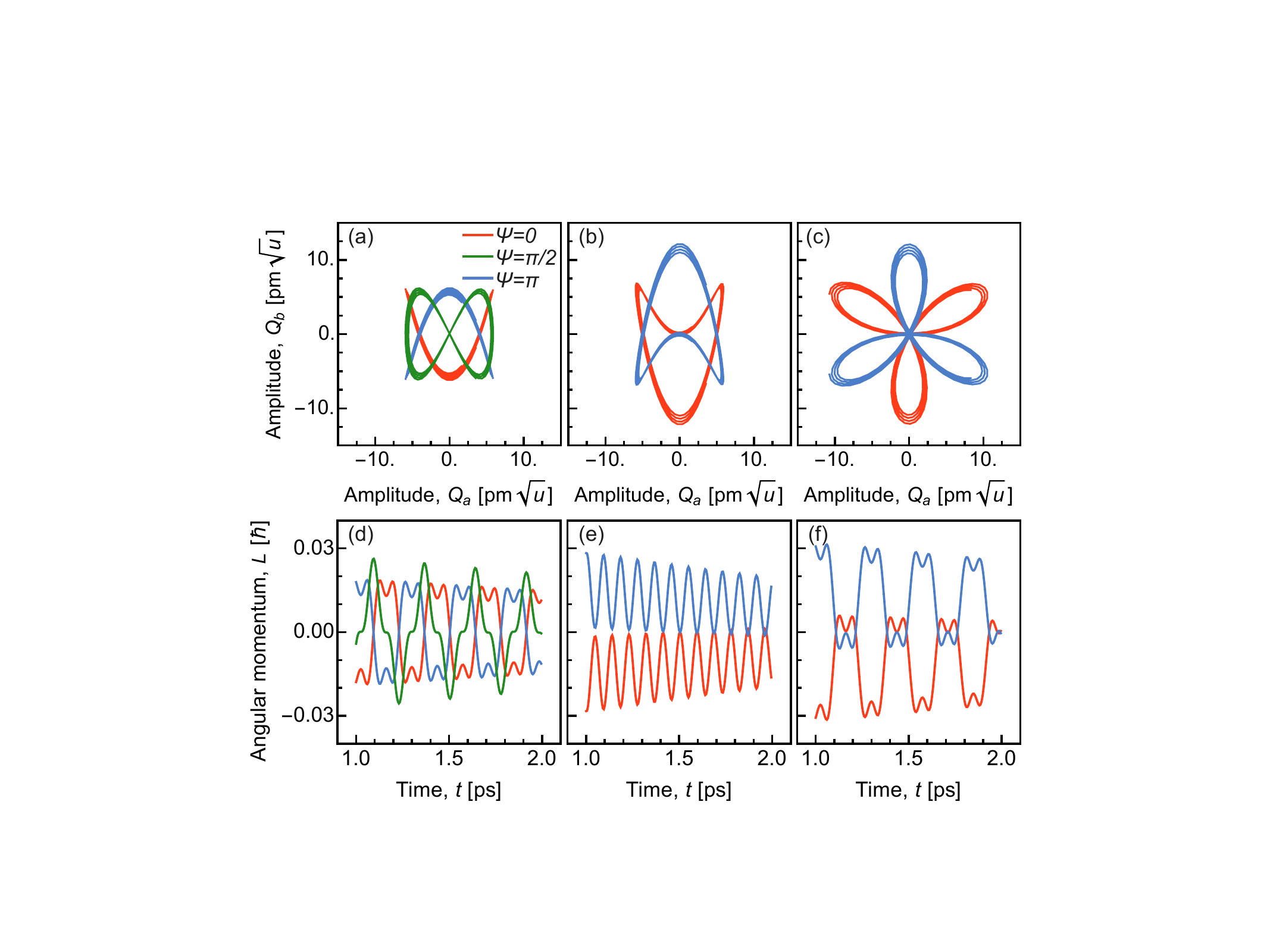}
\caption{
\label{fig:phase}
{\small
\textbf{Switching phononic Lissajous figures through phase control.} 
}
        Adding a $\pi$ phase shift to the carrier envelope phase of the $b$ component of the electric field reorients the direction of the resulting Lissajous figure. For each polarization, both (a--c) the Lissajous figure and (d--f) the corresponding phonon angular momentum are displayed for the time period of 1~ps to 2~ps after excitation. In the linearly polarized excitation, the eight-curve trajectory further deforms into a banana shape, which can be reversed.
}
\end{figure}

Having established a way to generate well-defined phononic Lissajous figures, we now turn to manipulating them in real space. The polarization configuration can be controlled through the carrier envelope phase, which allows us to reorient the phononic trajectories in real space. To demonstrate this, we introduce an additional phase $\psi$ added to the carrier envelope phase of the $b$ component of the laser pulse, $\phi_{\alpha,b}\rightarrow\phi_{\alpha,b}+\psi$, and solve the equations of motion Eq.~\eqref{eq:A1} and \eqref{eq:IR} for $\psi=0,\pi$. Fig.~\ref{fig:phase}(a), shows the phonon trajectories for the linear, mixed circular-linear, and circular polarizations (left to right, determined by the parameters shown in Table~\ref{tab:merged_polarization}), at the ideal $\omega:2\omega$ frequency ratio. When the additional $\pi$ phase is introduced, we observe an inversion of the Lissajous figures, allowing us to control the spatial symmetry breaking. Fig.~\ref{fig:phase}(b) further shows the phonon angular momentum corresponding to the trajectories in (a), which reverses its sign when the additional $\pi$ phase is introduced. For the linearly excited Lissajous figures, the phonon angular momentum oscillates and averages to zero in time. For the circularly-linearly and circularly excited Lissajous figures in contrast, nonzero net angular momentum is produced even after time averaging, similar to the case of purely circular motion of the atoms. Intriguingly, the banana-shaped trajectories under linear excitation and the cloverleaf trajectories under circular excitation produce angular momentum resembling a rectangular pulse train, possibly enabling new ways of angular momentum coupling in solids.

\begin{figure}[t]
\centering
\includegraphics[scale=0.4]{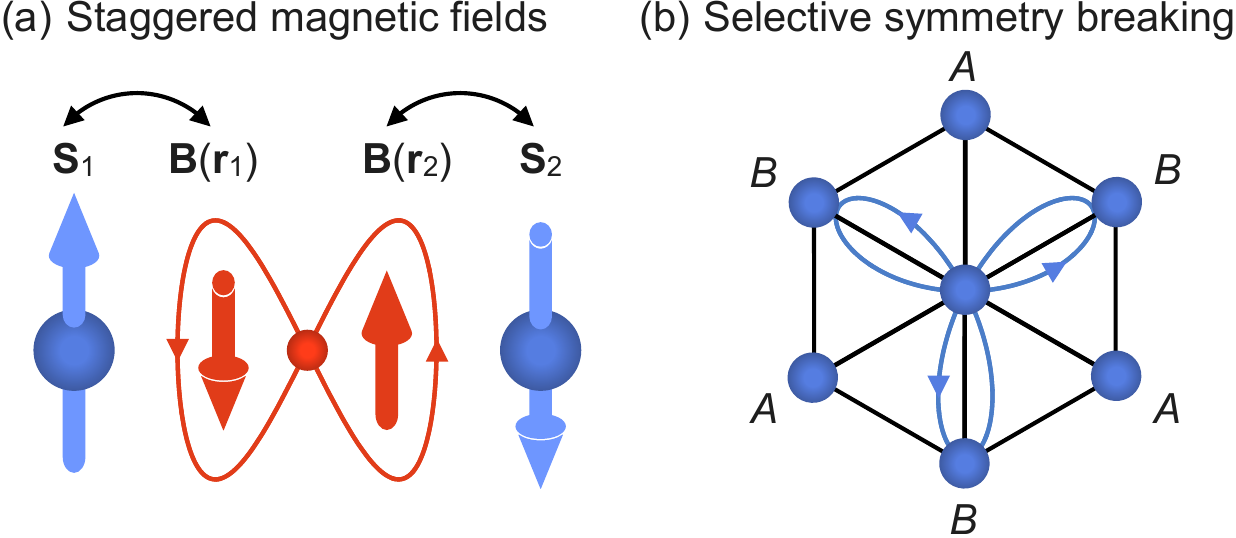}
\caption{
\label{fig:applications}
{\small
\textbf{Floquet-driving applications.} 
}
        (a) A phononic eight-curve trajectory leads to a staggered angular momentum and therefore effective magnetic field, $\mathbf{B}(\mathbf{r})$, which can potentially be used to couple to and switch antiferromagnetically ordered spins, $\mathbf{S}_1$ and $\mathbf{S}_2$. (b) A phononic cloverleaf trajectory can selectively change the bonding to neighboring atoms, as shown here for a triangular lattice in which nearest neighbors are split into two subsets, $A$ and $B$.
}
\end{figure}


\section{Discussion}

Our method of generating multicolor phonons in terahertz cavities enables phonon angular momentum shaping and at the same time opens a route towards advanced phononic Floquet driving to engineer the properties of quantum materials. We provide two examples of how our technique could be used in Fig.~\ref{fig:applications}. First, the generation of spatially separated angular momentum, as in the eight-curve and cloverleaf trajectories, will produce staggered effective magnetic fields that could be used to control and switch antiferromagnetically aligned spins (Fig.~\ref{fig:applications}(a)), which is not possible with uniform phonomagnetic fields produced by conventional circularly polarized phonons \cite{Luo2023,Basini2024,Davies2024}. The magnitude of the phonon magnetic moment is strongly material dependent and ranges from fractions of a nuclear magneton to several Bohr magnetons \cite{Juraschek2019,Geilhufe2021,Ren2021,Juraschek2022_giantphonomag,Saparov2022,Geilhufe2023,Zhang2023_BLG,Bonini2023,Shabala2024,Klebl2024,Chaudhary2024,Chen2025_gaugetheory}. Second, the spatial symmetry breaking induced by the phononic Lissajous trajectories allows for bond-selective changes of interactions in the solid. We schematically illustrate the example of a triangular lattice in Fig.~\ref{fig:applications}(b), where the cloverleaf trajectory breaks the nearest neighbors into two subsets, $A$ and $B$, possibly enabling the engineering of magnetic frustration. These and other applications will be studied in future work.

While previous studies of optical phonons in cavities focused on nonlinear effects \cite{Juraschek2021_4,OjedaCollado2024,Bostrom2024}, all mechanisms discussed in this work occur even in the harmonic approximation of phonons. We anticipate that developments in the field will accelerate in the next years, as terahertz cavities are becoming more widely applied and new cavity designs are developed \cite{Hubener2021,Schlawin2022}.

\begin{acknowledgments}
We thank Benoit Truc, Gregor Jotzu, Michael Fechner, Hannes H\"{u}bener, and Ofer Neufeld for useful discussions. This work was supported by the Israel Science Foundation (ISF) Grant No. 1077/23 and 1916/23. D.M.J. acknowledges support from the ERC Starting Grant CHIRALPHONONICS, no. 101166037.

\end{acknowledgments}



%


\onecolumngrid
\clearpage

\setcounter{page}{1}

\begin{center}
\textbf{\large Supplemental Material:\\ Multicolor phonon excitation in terahertz cavities}\\[0.4cm]
Omer Yaniv$^{1}$ and Dominik M. Juraschek$^{1,2}$ \\[0.15cm]
$^1${\itshape{\small School of Physics and Astronomy, Tel Aviv University, Tel Aviv 6997801, Israel}}\\
$^2${\itshape{\small Department of Applied Physics and Science Education,\\ Eindhoven University of Technology, 5612 AP Eindhoven, Netherlands}}\\
\end{center}

\setcounter{equation}{0}
\setcounter{figure}{0}
\setcounter{table}{0}
\setcounter{section}{0}
\makeatletter
\renewcommand{\theequation}{S\arabic{equation}}
\renewcommand{\thefigure}{S\arabic{figure}}
\renewcommand{\thetable}{S\arabic{table}}

\section{View of Phonon Eigenvectors Along the $c$ axis}
\begin{figure*}[h]
    \centering
    \includegraphics[scale=0.4]{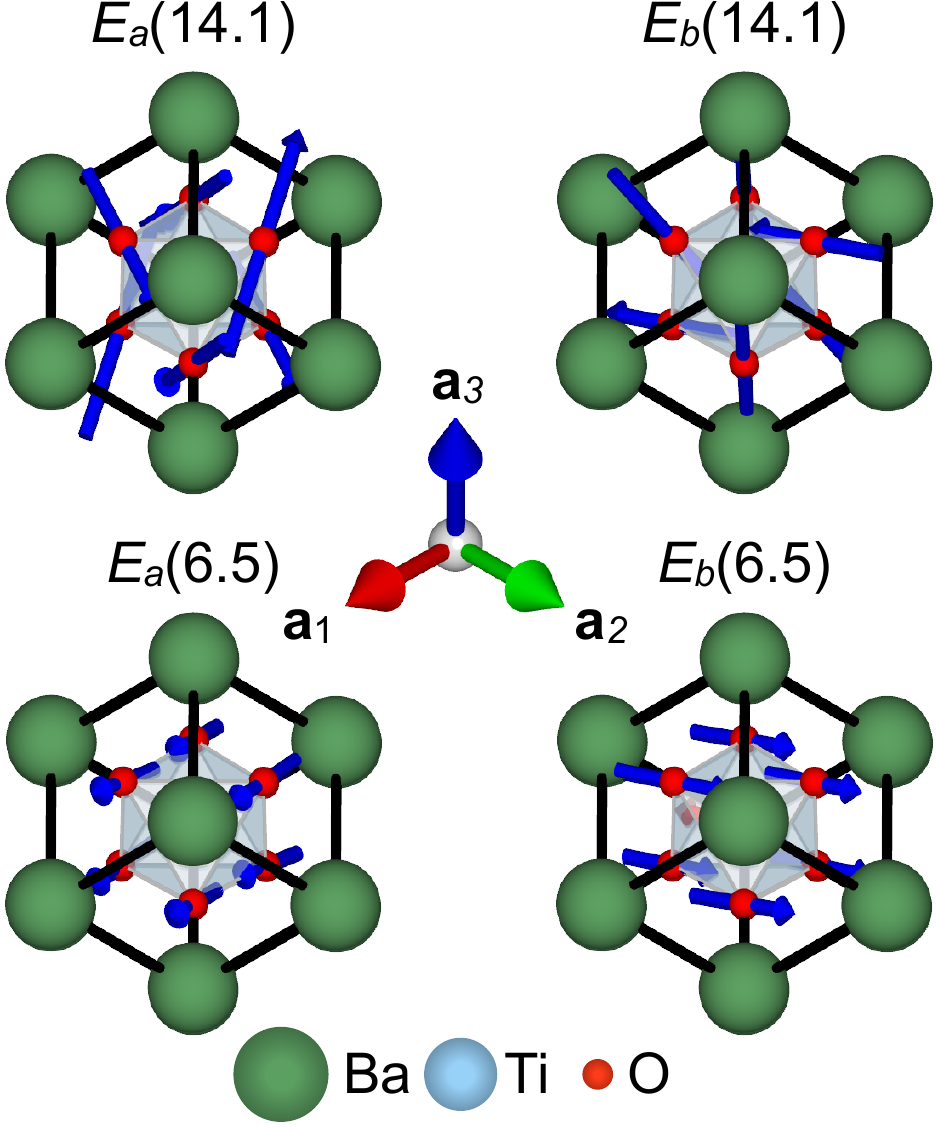}
    \caption{
        \small
        \textbf{Phonon eigenvectors along the [111] direction.} 
        Atomic displacements for the orthogonal components of the doubly degenerate $E$(14.1) and $E$(6.5) modes in the rhombohedral phase of BaTiO$_3$. The crystal is viewed along the $c$ axis, corresponding to the [111] direction. $\mathbf{a}_1$, $\mathbf{a}_2$, and $\mathbf{a}_3$ denote the primitive eigenvectors of the unit cell.
    }
    \label{fig:phonon_eigenvectors}
\end{figure*}

To further illustrate the phonon displacement patterns discussed in the main text, we present an alternative perspective of the atomic eigenvectors in Fig.~\ref{fig:phonon_eigenvectors}, viewed along the $c$ axis, corresponding to the [111] direction, of the crystal.
    
\section{Cavity-Phonon Polariton Branches}

The cavity-phonon polaritons correspond to the eigenmodes of the coupled equations of motion, Eqs.~\eqref{eq:A1} and \eqref{eq:IR} in the main text. We obtain the polariton branches as a function of the fundamental cavity frequency by seeking solutions of the form
\begin{equation}
A(t) = A_0\,e^{\,i\Omega_p t},  
\quad  
Q_\nu(t) = Q_0\,e^{\,i\Omega_p t}.
\end{equation} 
This ansatz assumes harmonic oscillations of both the cavity and the phonon mode with frequency \(\Omega_p\), allowing us to transform the equations of motion into frequency space. Substituting these expressions into Eqs.~\eqref{eq:A1} and \eqref{eq:IR} from the main text yields
\begin{align}
(\omega_c^2 - \Omega_p^2 + i\,\kappa_c\Omega_p)\,A_0 + D\,\Omega_p^2\,Q_0 &= 0, \\
(\Omega_\nu^2 - \Omega_p^2 + i\,\kappa_{\nu}\Omega_p)\,Q_0 - G\,A_0 &= 0.
\end{align}  
This system can be rewritten in matrix form as  
\begin{equation}
\begin{pmatrix}
\omega_c^2 - \Omega_p^2 + i\,\kappa_c\Omega_p & D\,\Omega_p^2 \\[6pt]
-\,G & \Omega_{\nu}^2 - \Omega_p^2 + i\,\kappa_{\nu}\Omega_p
\end{pmatrix}
\begin{pmatrix}
A_0\\[6pt]
Q_0
\end{pmatrix}
=
\begin{pmatrix}
0\\[6pt]
0
\end{pmatrix}.
\end{equation}
For nontrivial solutions to exist, the determinant of the coefficient matrix must vanish. Setting the determinant to zero yields the characteristic equation  
\begin{equation}
\bigl(\omega_c^2 - \Omega_p^2 + i\,\kappa_c\Omega_p\bigr)\bigl(\Omega_\nu^2 - \Omega_p^2 + i\,\kappa_\nu\Omega_p\bigr)
+GD\,\Omega_p^2 = 0.
\end{equation}  
By solving this equation numerically, we determine the polariton frequencies, corresponding to the real parts of \(\Omega_p\), which describe the hybridized cavity-phonon polariton branches as a function of the fundamental cavity frequency.

\section{Multicolor phonon dynamics for branches 3 and 4}

\begin{figure*}[h]
    \centering
    \includegraphics[scale=0.515]{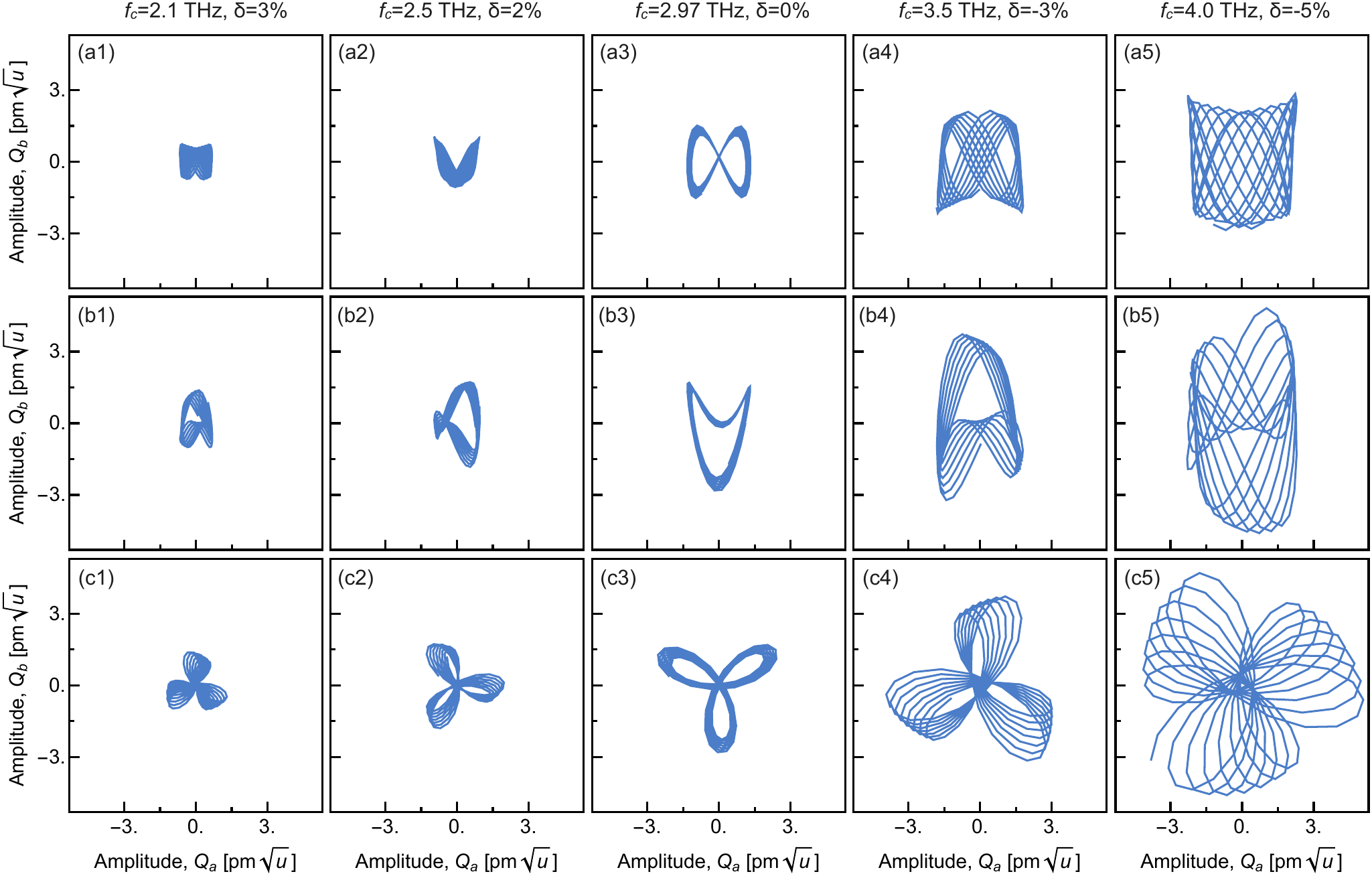}
    \caption{
        \small
        \textbf{Phononic Lissajous figures arising from branches 4 and 3.} Time evolution of the phonon trajectories in the $ab$ plane of the crystal for the time interval of 1~ps to 2~ps after excitation. We show the dynamics for five different cavity frequencies, tuning the cavity-phonon polariton frequencies into and out of the ideal $\omega:2\omega$ frequency ratio, reached at $\omega_c=2.97$~THz and yielding polariton frequencies of $\Omega_4=14.17$~THz and $\Omega_3=7.09$~THz.
        When the cavity frequency satisfies the  $\omega:2\omega$  ratio, the Lissajous figures exhibit closed trajectories, leading to steady-state crystallographic symmetry breaking. As the cavity frequency is detuned, the phononic trajectories progressively deviate from closed orbits, disrupting the steady-state symmetry breaking.
        (a1-a5) Dynamics for linear excitations of the polariton branches, creating an eight curve when reaching the ideal frequency ratio. (b1-b5) Dynamics for mixed circular-linear  excitations, creating an arrowhead curve when reaching the ideal frequency ratio. (c1-c5) Dynamics for circular excitations, creating a cloverleaf curve when reaching the ideal frequency ratio.}
    
    \label{fig:phononic_lissajous}
\end{figure*}

Here, we present the trajectories of the phononic Lissajous figures arising from the excitation of cavity-phonon polariton branches 4 and 3, as referenced in the main text. Branch 4 corresponds to the upper polariton of the $E(14.1)$ mode and branch 3 corresponds to the upper polariton of the $E(6.5)$ mode. At a cavity frequency of $\omega_c=2.97$~THz, the cavity-phonon polaritons lie at frequencies $\Omega_4=14.17$~THz and $\Omega_3=7.09$~THz. By numerically solving Eqs.~\eqref{eq:A1} and \eqref{eq:IR} in the main text for varying cavity frequencies around 2.97~THz, we illustrate how these excitations can be tuned to form phononic Lissajous figures in Fig.~\ref{fig:phononic_lissajous}.

\end{document}